\documentclass[12pt]{article}
\usepackage{latexsym}
\usepackage{amsmath,,calrsfs}
\usepackage{amsfonts}
\usepackage{amssymb}
\usepackage{amscd}
\usepackage{fancybox}
\usepackage{cite}
\usepackage{amsmath,amsfonts,amsbsy}
\usepackage{pstricks,pst-node}
\usepackage[small,bf,hang]{caption2}

\usepackage{float}

\psset{unit=1.3cm,linewidth=.5pt,radius=.2}  % controls the size of diagrams

\def\hybrid{\topmargin -20pt    \oddsidemargin 0pt
        \headheight 0pt \headsep 0pt
        \textwidth 6.25in       % A4 paper
        \textheight 9.5in       % A4 paper
        \marginparwidth .875in
        \parskip 5pt plus 1pt   \jot = 1.5ex}

\hybrid

\newcommand{\beq}{\begin{equation}}
\newcommand{\eeq}{\end{equation}}
\newcommand{\bi}{\begin{itemize}}
\newcommand{\ei}{\end{itemize}}
\newcommand{\bt}{\begin{tabular}}
\newcommand{\et}{\end{tabular}}
\newcommand{\bc}{\begin{center}}
\newcommand{\ec}{\end{center}}

\newcommand{\be}{\begin{equation}}
\newcommand{\ee}{\end{equation}}
\newcommand{\bea}{\begin{eqnarray}}
\newcommand{\eea}{\end{eqnarray}}
\newcommand{\ba}{\begin{array}}
\newcommand{\ea}{\end{array}}

\def\bbox{{\,\lower0.9pt\vbox{\hrule \hbox{\vrule height 0.2 cm
\hskip 0.2 cm \vrule height 0.2 cm}\hrule}\,}}
\newcommand{\dsl}{\pa \kern-0.5em /}

\begin{document}

\begin{titlepage}%1
\begin{center}

\hfill UG-09-75 \\
\hfill MIT-CTP-4085 \\
\hfill DAMTP-2009-77

\vskip 2cm

{\Large \bf   On Higher Derivatives in  3D Gravity and \\ 
\bigskip
Higher Spin Gauge Theories}

\vskip 1.5cm

{\bf Eric A.~Bergshoeff\,$^1$, Olaf Hohm\,$^2$ and Paul K.~Townsend\,$^3$} \\

\vskip 30pt

{\em $^1$ \hskip -.1truecm Centre for Theoretical Physics, University of Groningen, \\
Nijenborgh 4, 9747 AG Groningen, The Netherlands \vskip 5pt }

{email: {\tt E.A.Bergshoeff@rug.nl}} \\

\vskip 15pt

{\em $^2$ \hskip -.1truecm Centre for Theoretical Physics,\\ Massachusetts Institute of Technology,\\
Cambridge, MA 02139, USA \vskip 5pt }

{email: {\tt ohohm@mit.edu}} \\

\vskip 15pt

{\em $^3$ \hskip -.1truecm Department of Applied Mathematics and
Theoretical Physics,\\
Centre for Mathematical Sciences, University of Cambridge,\\
Wilberforce Road, Cambridge, CB3 0WA, U.K. \vskip 5pt }

{email: {\tt P.K.Townsend@damtp.cam.ac.uk}} \\

\end{center}

\vskip 1cm

\begin{center} {\bf ABSTRACT}\\[3ex]
\end{center}

The general second-order massive field equations for arbitrary positive integer spin in three spacetime dimensions, and their  
``self-dual''  limit to  first-order equations,  are shown to be equivalent to  gauge-invariant higher-derivative field equations.  We recover  most known  equivalences for spins $1$ and $2$, and find some new ones.  In particular, we find a non-unitary  massive 3D gravity theory with  a 5th order term obtained by contraction of the Ricci and Cotton tensors; this term is part of an ${\cal N}=2$ super-invariant that includes the ``extended Chern-Simons'' term of 3D electrodynamics.   We also find a new {\it unitary}   6th order  gauge theory  for   ``self-dual''  spin $3$.

\begin{minipage}{13cm}
\small

\end{minipage}

%\today

%\end{center}

%\noindent

\vfill

%\July 2008

\end{titlepage}

\newpage
\setcounter{page}{1}
\pagestyle{plain}
\tableofcontents

%\newpage

\section{Introduction}\setcounter{equation}{0}

A peculiar feature of relativistic  gauge field theory in three spacetime dimensions (3D) is that the particles associated with the gauge field may be massive even if the gauge symmetry is unbroken. In the  well-known examples, such as ``topologically massive electrodynamics''  (TME) \cite{Siegel:1979fr,Schonfeld:1980kb,Deser:1982vy,Gates:1983nr} or ``topologically massive gravity'' (TMG) \cite{Deser:1981wh},  the non-zero mass arises because of some parity-breaking topological term in the action, but the recent example of ``new massive gravity'' (NMG) \cite{Bergshoeff:2009hq} demonstrates that the phenomenon is more general than had previously been thought; NMG  is a generally covariant and parity preserving theory for interacting massive spin 2 particles with two polarization states of helicities 
$\pm2$. An obvious question is whether there are other massive 3D gauge theories waiting to be discovered, and whether there is some systematic way of finding them. The main purpose of this paper is to show how  gauge theories describing massive particles of integer spin may be found in a systematic way starting from standard second-order (or sometimes first-order) free field equations. Along the way we will recover a number of   known gauge theory models, but we will also find a few new ones of  interest.  

Let us first recall the Fierz-Pauli (FP) field equations \cite{Fierz-Pauli}  for a symmetric rank-$s$ tensor field describing  massive particles of integer spin $s$ on a four-dimensional Minkowski spacetime (with `mostly plus' signature): 
\be\label{FPeqs}
\left(\Box -m^2\right) \phi_{\mu_1\cdots \mu_s} =0\, , \qquad \partial^\mu\phi_{\mu\nu_1\cdots \nu_{s-1}} =0\, , \qquad 
\eta^{\mu\nu} \phi_{\mu\nu\rho_1\cdots \rho_{s-2}} =0\, . 
\ee
The differential subsidiary condition is needed for positivity of the energy (and hence unitarity) and the tracelessness condition is needed to avoid the propagation of lower spin modes; see e.g. \cite{Bouatta:2004kk}.  The same equations may be used if the Minkowski spacetime is 3-dimensional, and in this case they may be rewritten as
\be\label{3DFP2}
\left[ {\cal P}(m) {\cal P}(-m)\right]_{\mu_1}{}^\rho  \phi_{\rho\mu_2\cdots\mu_s} =0 \, , 
\qquad \eta^{\mu\nu}\phi_{\mu\nu\rho_1\cdots\rho_{s-2}} =0\, , 
\ee
where the operator
\be
{\cal P}(m)_\mu{}^\nu = \frac{1}{2} \left[\delta_\mu{}^\nu - \frac{1}{m} \varepsilon_\mu{}^{\tau\nu}\partial_\tau\right]
\ee
is an on-shell projection operator:
\be
{\cal P}^2(m) \phi =  {\cal P}(m) \phi  \qquad {\rm if} \  \phi  \ {\rm satisfies} \ (\ref{3DFP2}). 
\ee
Note that the equations (\ref{3DFP2}) imply the differential subsidiary condition, and  that the tensor ${\cal P}(m) {\cal P}(-m)\phi$ is symmetric on its $s$ indices as a consequence of the subsidiary conditions. The operator ${\cal P}$ projects onto modes of helicity $\pm1$ within the space of divergence-free vector fields satisfying the mass $m$ wave equation (the sign depending on the sign of $m$)  and hence onto modes of helicity $\pm s$ when acting on the rank-$s$ symmetric tensor $\phi$. The equations (\ref{3DFP2}) therefore propagate  one mode of helicity $s$ and another of helicity $-s$, both with mass $m$. These are the two helicity states of a massive particle of spin $s$ in three spacetime dimensions. Actually, each helicity state yields an irreducible induced representation of the  3D Poincar\'e group but a parity preserving theory of  spin $s$ must propagate both helicities $\pm s$ with the same mass because
parity flips the sign of the helicity.

There is a  generalization of the 3D FP equations  (\ref{FPeqs})  that  breaks parity. Consider the following equations 
for  independent positive masses $m_\pm$:
\be\label{genFP}
\left[ {\cal P}(m_+) {\cal P}(-m_-)\right]_{\mu_1}{}^\rho  \phi_{\rho\mu_2\cdots\mu_s} =0 \, , 
\qquad \eta^{\mu\nu}\phi_{\mu\nu\rho_1\cdots\rho_{s-2}} =0\, . 
\ee
These equations propagate one mode of helicity $s$ with mass $m_+$ and another of helicity $-s$ with mass $m_-$; we shall refer to them as the ``generalized  (3D) Fierz-Pauli'' equations.  This generalization changes the  {\it dynamical}  equation from the mass $m$ wave equation to the new equation 
\be\label{newwave}
\left(\Box -m^2\right) \phi_{\mu_1\cdots \mu_s} =\tilde\mu\,  \varepsilon_{\mu_1}{}^{\tau\nu} \partial_\tau \phi_{\nu\mu_2\cdots\mu_s}\, , 
\ee
where 
\be
m^2= m_+m_- \, , \qquad \tilde \mu = m_- - m_+\, .  
\ee
The subsidiary conditions are unchanged (as long as $m^2\ne0$) and these imply that  the right hand side of (\ref{newwave}) is symmetric in its $s$ free indices.  We may assume without loss of generality that $\tilde\mu\ge0$, and take the limit that $\tilde\mu \to\infty$ for fixed
\be
\mu \equiv m^2/\tilde\mu\, , 
\ee
which implies that $m^2\to\infty$. Equivalently, we take $m_-\to\infty$ for fixed $m_+$. In this limit the helicity $-s$ mode decouples and we have a single mode of helicity $s$ and mass $\mu$ described by the first-order equation
\be
{\cal P}(\mu)_{\mu_1}{}^\rho\phi_{\rho\mu_2 \cdots \mu_s} =0 \, , \qquad \eta^{\mu\nu}\phi_{\mu\nu\rho_1\cdots\rho_{s-2}} =0\, . 
\ee
The differential subsidiary condition is again implied. Following the terminology introduced for the spin 1 case in  \cite{Townsend:1983xs} and subsequently used for the spin 2 case in \cite{Aragone:1986hm}, we shall call this the ``self-dual'' theory of a spin $s$ particle of mass $\mu$ in 3D; the arbitrary spin case was studied in \cite{Tyutin:1997yn}. 

We will show how these  various sets of equations for a rank-$s$ tensor $\phi$ may be replaced, in two different ways when $s\ge2$,  by equations for a rank-$s$ gauge {\it potential}. The method was first used for $s=2$ in \cite{Andringa:2009yc} and we will shortly summarize the results obtained there. It also applies for $s=1$, in which case  the FP equations become the Proca equations; we shall use the notation $B$ in place of $\phi$ for the Proca vector field.   Although  this case is too simple to fully illustrate the method, it is a good starting point. Specifically, let us 
start from the  self-dual spin 1  model,  for which the (unitary) action is  
\be\label{sqrtPaction}
S_{SD}[B] =\int\,d^3 x\,\left\{  \tfrac{1}{2} \varepsilon^{\mu\nu\rho} B_\mu\partial_\nu B_\rho +  \tfrac{1}{2}  \mu\, B^2 \right\}\, . 
\ee
One may easily verify that the field equation is 
\be\label{Om}
{\cal P}(\mu)_\mu{}^\nu B_\nu=0\, ,  
\ee
and that this implies the differential subsidiary condition $\partial\cdot B =0$, which has the following general solution in terms of 
a gauge potential $A$, defined up to a Maxwell gauge transformation: 
\be\label{BtoA}
B^\mu  = \tfrac{1}{2}\varepsilon^{\mu\nu\rho} F_{\nu\rho} \equiv \tilde F^\mu\, , \qquad F_{\mu\nu} \equiv  2\partial_{[\mu}A_{\nu]}\, .
\ee
When this  is used in (\ref{Om}), we find the new equation\footnote{It  is not generally permissible to substitute the solution of a field equation into the action, and no such substitution will be contemplated in this paper.}
\be
{\cal P}(\mu)_\mu{}^\nu\, \tilde F_\nu = 0\, ,
\ee
which is the equation of motion of  TME:
\be
S_{TME}[A] = \int\,d^3 x\,\left\{ \tfrac{1}{2}\tilde F^2   + \tfrac{1}{2} \mu\,\varepsilon^{\mu\nu\rho} A_\mu\partial_\nu A_\rho\right\}\, .
\ee
This  establishes on-shell equivalence of the self-dual and TME theories; i.e. the equivalence of their field equations.  Off-shell equivalence then follows because we are free to choose the sign (and have done so) such that the TME action is also unitary; in other words, the one propagated mode is physical, rather than a ghost.

We should stress that the equivalence of the self-dual  and TME theories is not a new result; the off-shell equivalence was proved in  \cite{Deser:1984kw} by means of a ``master action''.  It was also observed in  \cite{Deser:1984kw} that  the equations of TME are those of the self-dual model with $A\to \tilde F(A)$, but  the on-shell equivalence does not follow from this fact alone because the same replacement  applied to TME yields the {\it inequivalent}  equations of   ``extended topologically massive electrodynamics'' (ETME) 
\cite{Deser:1999pa}, as we review in section {\ref{sec:spin1}.  In contrast, the  procedure of solving subsidiary conditions manifestly yields equivalent equations, although there is no guarantee that the corresponding {\it actions} are equivalent because it may be that 
the new action is not unitary.  This point is well illustrated by the  the 3D Proca theory for a spin 1 particle of mass $m$:  the field equations are
\be
\left(\square -m^2\right) B_\mu =0\, , \qquad \partial\cdot B =0\, .
\ee
Solving the subsidiary condition as before, we arrive at the ``extended Proca'' (EP) equation $\left(\square-m^2\right)\tilde F_\mu=0$, which is derivable from the following action containing the ``extended Chern-Simons'' (ECS) term introduced in \cite{Deser:1999pa}
\be\label{EPact}
S_{EP}[A] = \int\,d^3 x\,\left\{ \tfrac{1}{2}\varepsilon^{\mu\nu\rho} \tilde F_\mu\partial_\nu \tilde F_\rho -
\tfrac{1}{2}m^2 \varepsilon^{\mu\nu\rho} A_\mu\partial_\nu A_\rho \right\}\, .
\ee
By construction,  this model is on-shell equivalent to Proca,  and this implies the propagation of two modes of mass $m$, one with helicity $+1$ and the other with helicity $-1$. However,  the on-shell equivalence to Proca does {\it not} extend to an off-shell equivalence because one of the two helicity modes propagated by the EP  theory is a ghost, as we show later. We believe that this  clarifies the very  brief discussions of this model in \cite{Deser:1999pa,Deser:2008rm}, where it was merely observed that  it leads to 
``massive propagation of the field strength''.

Moving on to spin 2, let us apply the analogous `trick'  to the self-dual spin 2 model with  field equations
\be\label{SDspin2}
{\cal P}(\mu)_\mu{}^\rho \phi_{\rho\nu} =0\, , \qquad  \eta^{\mu\nu} \phi_{\mu\nu} =0\, . 
\ee
In other words, we solve the differential subsidiary condition $\partial^\mu\phi_{\mu\nu}=0$ that these equations imply. The general solution involves a second-order differential operator because the symmetry of $\phi_{\mu\nu}$ means that it is divergence-free in both indices if it is divergence-free in one (as the subsidiary condition says it is). In fact, the general solution is 
\be\label{tildehG}
\phi_{\mu\nu} = {\cal G}_{\mu\nu}{}^{\rho\sigma} h_{\rho\sigma} \equiv  G^{\rm (lin)}_{\mu\nu}\, , \qquad 
{\cal G}_{\mu\nu}{}^{\rho\sigma} \equiv -\tfrac{1}{2} \varepsilon_{(\mu} {}^{\eta\rho}\varepsilon_{\nu)} {}^{\tau\sigma}\partial_\eta\partial_\tau  
\ee
for some symmetric tensor potential $h_{\mu\nu}$.  The matrix operator ${\cal G}$ is  the ``Einstein operator'' in the form introduced in \cite{Bergshoeff:2009hq}; the terminology comes from the fact that  $G^{\rm (lin)}_{\mu\nu}$ is the linearization of the Einstein tensor for the metric $g_{\mu\nu}=\eta_{\mu\nu} + h_{\mu\nu}$.   The  self-dual  field equation of (\ref{SDspin2}) now becomes
\be\label{3rdorder}
{\cal P}(\mu)_\mu{}^\rho G^{\rm (lin)}_{\rho\nu} =0\, , \qquad   \eta^{\mu\nu} G^{\rm (lin)}_{\mu\nu} =0\, , 
\ee
but the second of these is implied by the first, which is the linearized limit of the field equation of ``topologically massive gravity'' (TMG):
\be\label{TMGeq}
G_{\mu\nu} + \frac{1}{\mu} C_{\mu\nu} =0\;, \,  \qquad \sqrt{|g|}\, C_{\mu\nu}\equiv \varepsilon_{(\mu}{}^{\tau\rho}D_{|\tau} G_{\rho | \nu)}\, , 
\ee
where $|g|= -\det g$ for metric tensor $g$. 
The tensor $G$ is the Einstein tensor and $C$ is the Cotton tensor,  the 3D analog of the 4D Weyl tensor. The TMG action is the sum 
of  a Lorentz Chern-Simons (LCS) term, the variation of which is proportional to the Cotton tensor, and a `wrong sign'  Einstein-Hilbert (EH) term (i.e. with the sign opposite to the usual one; the sign itself is convention dependent). TMG is unitary despite the fact that the LCS term is 3rd order in derivatives because one may choose the overall sign of the action such that the one propagating mode is physical, rather than a ghost; this  fixes the sign of the EH term. We should again stress that the equivalence of linearized TMG to the self-dual spin 2 theory  is not a new result. It was proved by a `master action' method in \cite{Aragone:1986hm}, where the first-order self-dual spin 2 model was introduced\footnote{Linearized TMG was also proved to be equivalent to a second-order  self-dual model in \cite{Deser:1990ay}; this action is the  $m^2\rightarrow 0$ limit of the action (\ref{eact}) to be considered below. The
equivalence of all three spin 2 models was made manifest  in  \cite{Dalmazi:2008zh},  and a simplified version of this `triple' 
equivalence was presented in  \cite{Bergshoeff:2009hq}. For an earlier overview of spin-$2$ equivalences, with an emphasis on implications for interactions, see \cite{Deser:1992md}.}.

We can repeat the above procedure  for the  FP spin 2 theory itself, and its `generalized' version,  and this was done in  \cite{Andringa:2009yc}.  The resulting 4th order equations are the linearized equations  of  ``new massive gravity''  (NMG) and its extension to ``generalized massive gravity'' (GMG)  \cite{Bergshoeff:2009hq}.  NMG has the action 
\be
I_{NMG} \propto \int d^3x \, \sqrt{|g|} \left[  R - \frac{1}{m^2} K \right] \, , \qquad K \equiv R_{\mu\nu}R^{\mu\nu} - \frac{3}{8} R^2\, , 
\ee
where $R_{\mu\nu}$ is the Ricci tensor and $R$  the Ricci scalar.  GMG has, additionally, an LCS term with coefficient $1/\mu$, so one recovers NMG in the limit $\mu^2\to\infty$ for fixed $m$ and TMG in the limit $m^2\to\infty$ for fixed $\mu$.  For NMG, this proof of linearized equivalence with FP is closely analogous to  the  proof  presented above that  extended Proca is  on-shell equivalent  to Proca. That analogy,  and the fact that we now have 4th order field equations, might lead one to expect  one of the two spin 2 modes of NMG to be a ghost. Remarkably however, NMG is unitary. This was originally shown by an  auxiliary field method that allows one to pass directly from the linearized  NMG action to the unitary  FP action \cite{Bergshoeff:2009hq,Nakasone:2009bn}; the absence of ghosts was subsequently confirmed  by a canonical analysis \cite{Deser:2009hb}. The auxiliary field method does not seem to be as useful in  the GMG case but the absence of ghosts in this case was demonstrated by canonical methods in  \cite{Andringa:2009yc}.

The method that we have now illustrated for spin 1 and spin 2 can be extended further in the spin 2 case \cite{Andringa:2009yc}:  we may solve {\it both} the differential subsidiary condition and the algebraic tracelessness condition simultaneously by writing 
\be\label{newsub}
\phi_{\mu\nu} = \varepsilon_{(\mu}{}^{\tau\lambda} \partial_{|\tau} {\cal G}_{\lambda | \nu)}{}^{ \rho\sigma} h_{\rho\sigma} \equiv 
C_{\mu\nu}^{\rm (lin)} \, , 
\ee
where the symmetric tensor potential $h$ is now defined up to a linearized diffeomorphism {\it and} a linearized Weyl transformation. 
If this is used in the self-dual equations (\ref{SDspin2}), one finds 4th order equations  that are the linearization of the equations 
obtained from GMG by omitting the EH term. This model was called ``New Topologically Massive Gravity'' (NTMG) in \cite{Andringa:2009yc}; by construction, it is equivalent to the spin 2 self-dual model, as a canonical analysis confirms. The same model was studied independently in \cite{Dalmazi:2009pm}. 

Solving {\it both}  subsidiary conditions for the generalized FP theory yields 5th order equations that also have an obvious non-linear generalization; the 5th order term in the action is obtained from the contraction of the Ricci tensor with  the Cotton tensor.  Another purpose of this paper is to investigate  this new spin 2 field theory.  Let us recall here that the Cotton tensor $C_{\mu\nu}$ can be  be written in terms of the 3D Schouten tensor $S_{\mu\nu}$ as follows
\be\label{CotSchout}
C_{\mu\nu} \equiv  \frac{1}{\sqrt{|g|}} \varepsilon_\mu{}^{\tau\rho} D_\tau S_{\rho\nu}\, , \qquad
S_{\mu\nu} \equiv R_{\mu\nu} - \tfrac{1}{4} g_{\mu\nu} R\, .
\ee
Thus, we find ourselves having to consider  a contribution to the Lagrangian density of the form\footnote{We understand that models including this term have been investigated in unpublished work of S. de Haro and R. Jackiw.} 
\be\label{ELCSnonlin}
{\cal L}_{ELCS} = \sqrt{|g|}\, R^{\mu\nu} C_{\mu\nu} \equiv  \varepsilon^{\mu\nu\rho} S_\mu{}^\sigma D_\nu S_{\rho\sigma}\, .
\ee
The Schouten tensor  has an  interpretation as a gauge potential for ``conformal boosts'', so we see that the new 5th order term is a type of Chern-Simons term. We shall call it the ``extended  Lorentz Chern-Simons'' (ELCS) term because it is  analogous to the  ECS term of electrodynamics.  Indeed, we find that the ELCS term  leads to ghosts in exactly the same way as does the ECS term.

This result is nicely explained by the  ${\cal N}=2$ supersymmetric extension of 3D gravity models with an ELCS term.  The 
 ${\cal N}=2$ supersymmetric extension of  linearized GMG was presented in \cite{Andringa:2009yc}; it unifies this 4th order 3D gravity theory with the 2nd order theory for massive spin 1. Here we show that the inclusion of the ELCS term in the gravity sector implies the inclusion of the ECS term in the electrodynamics sector, because both are part of a single  ${\cal N}=2$ off-shell supersymmetric invariant. In addition, we obtain in this way a similar result for spin 3/2; one that is of course already implicit in the ${\cal N}=1$ supersymmetric extension: the inclusion of the ELCS term (\ref{ELCSnonlin}) implies the addition to the Lagrangian density 
 of a 4th order term of the form
\be\label{Cottinosquared}
{\cal L}=  \sqrt{|g|}\,  \bar {\cal C}^\mu {\cal C}_\mu\, ,
\ee
where ${\cal C}^\mu$ is the `gamma-traceless' Cottino vector-spinor  super-partner of the Cotton tensor  
\cite{Deser:1982sw,Gibbons:2008vi},  and $\bar{\cal C}^\mu$ is its Majorana conjugate (we suppress spinor indices); 
upon linearization \cite{Andringa:2009yc},
\be\label{Cottino}
{\cal C}_{\rm (lin)}^\mu = \gamma^\nu \partial_\nu {\cal R}_{\rm (lin)}^\mu + \varepsilon^{\mu\nu\rho} \partial_\nu {\cal R}^{\rm (lin)}_\rho  \, , \qquad
{\cal R}_{\rm (lin)}^\mu =  \varepsilon^{\mu\nu\rho} \partial_\nu \psi_\rho\, , 
\ee
where $\psi_\mu$ is the anti-commuting Majorana Rarita-Schwinger (RS) vector-spinor potential. Obviously, the term (\ref{Cottinosquared}) must lead to spinorial ghosts because they are needed to complete the on-shell ghost particle ${\cal N}=1$ supermultiplets implied by the existence of spin 2 ghosts.  

Although it is far from clear how one might introduce interactions consistent with higher-spin gauge invariances, it is of  interest to consider how the procedure that we have been outlining for spins 1 and 2 generalizes to higher spins, and one purpose of this paper  is to present some details of the application to spin 3. Obvious questions are whether there are analogs of the linearized TMG, NMG and NTMG actions  for spin 2. A spin-$3$ analog of TMG was proposed in \cite{Damour:1987vm} but this was based on the assumption (standard for 4D \cite{Curtright:1979uz}) that the symmetric-tensor parameter of the spin-$3$ gauge transformation should be traceless. This allows the construction of a 2nd order analog of the linearized Einstein tensor, and hence a 3rd order TMG-like action,  but this action was shown in  \cite{Damour:1987vm}  to propagate a spin-$1$ ghost in addition to the desired spin-$3$ mode. Our method leads naturally to a 3rd order analog of the Einstein tensor, and hence to a 4th order analog of TMG; although we present only the field equations of this topologically-massive spin-$3$ theory (the action likely requires auxiliary fields) we confirm that it propagates a single spin-$3$ mode.  There is also a natural spin-$3$ analog of linearized NMG but, as in the spin-$1$ EP case, it propagates one of the two spin-$3$ modes as a ghost.  Finally, we derive the spin-$3$ analog of linearized NTMG, off-shell as well as on-shell. This is a 6th order 
field theory that is both unitary, as we confirm by a canonical analysis, and propagates a single spin-$3$ mode. 

The organization of this paper is straightforward: we proceed sequentially through spins $s=1,2,3$.
There is not much left to say about spin 1, but we complete the job in section \ref{sec:spin1}. In section \ref{sec:spin2} we
focus on the new 5th order spin 2 theory, and we show how ${\cal N}=2$ supergravity combines the 3rd order ECS term of 
3D electrodynamics with the linearization of the 5th order ELCS term for spin 2. We then adapt the procedure to spin 3 in
section \ref{sec:spin3}; this involves the introduction of 3D analogs of the Einstein and Cotton tensors. We summarize in section
\ref{sec:discussion} and include a brief discussion of higher spins and other possible generalizations.

\section{Higher-derivative 3D electrodynamics}\label{sec:spin1} 
\setcounter{equation}{0}

We will now complete our discussion of spin 1 models by considering  the generalized Proca theory with action 
\be\label{genproca}
S[B] = \int \! d^3 x\, \left\{  \tfrac{1}{2}\tilde G^2 -  \tfrac{1}{2}\tilde \mu \, \varepsilon^{\mu\nu\rho} B_\mu\partial_\nu B_\rho -  \tfrac{1}{2}m^2 B^2\right\} \, ,  \qquad
\tilde G^\mu = \varepsilon^{\mu\nu\rho} \partial_\nu B_\rho\, . 
\ee
The field equation is 
\be
[{\cal P}(m_+){\cal P}(-m_-)]_\mu{}^\nu\, B_\nu=0\, , \qquad  m^2= m_+m_- \, , \qquad \tilde\mu = m_- - m_+ \ge0\, . 
\ee
This propagates the helicity $+1$ mode with mass $m_+$ and the helicity $-1$ mode with mass $m_-$. One may verify that the Hamiltonian density is positive definite, so these modes are propagated unitarily.  Taking $m_-\rightarrow\infty$ for fixed $m_+$ (or $\tilde\mu\to\infty$ for fixed $\mu=m^2/\tilde\mu$) yields the self-dual  model, while setting  $m_+=m_-$, i.e. $\tilde\mu=0$, leads to the parity-preserving Proca model. We have already seen how these special cases are equivalent to, respectively,  TME and the non-unitary EP theory of (\ref{EPact}).  We now consider the generic case. 

Solving the subsidiary condition  $\partial\cdot B=0$ as in (\ref{BtoA}),  we arrive at  the equation
\begin{equation}
[{\cal P}(m_+){\cal P}(-m_-)]_\mu{}^\nu\,\tilde F_\nu=0\, , \qquad \tilde F^\mu = \varepsilon^{\mu\nu\rho}\partial_\nu A_\rho\, . 
\end{equation}
This  equation follows from the action
\be\label{3term}
S[A] = \frac{1}{2}\int \! d^3 x \left\{- \varepsilon^{\mu\nu\rho} \tilde F_\mu \partial_\nu \tilde F_\rho  + 
 \tilde \mu \, \tilde F^2 + m^2 \varepsilon^{\mu\nu\rho} A_\mu \partial_\nu A_\rho  
 \right\}\, . 
\ee
This is the generic  ``three-term''  electrodynamics model considered in \cite{Deser:1999pa}. We have now established the  {\it on-shell} equivalence of this model to (\ref{genproca}) and this gives us the helicity content, which was not worked out  in   \cite{Deser:1999pa}.  However, there is no guarantee of off-shell equivalence. 

Following \cite{Deser:1999pa}, but choosing the gauge $\partial_iA_i=0$,  we may decompose  the vector potential as follows:
\be
A_0= \frac{1}{\sqrt{-\nabla^2}}\psi \, , \qquad A_i = \frac{1}{\sqrt{-\nabla^2}} \, \hat \partial_i \phi \, , \qquad 
\hat\partial_i \equiv \varepsilon^{ij} \partial_j \, .
\ee
The action (\ref{3term}) now reduces to
\be
S[\psi,\phi] = \int \! d^3 x \left\{ -  \psi\square\phi +  \frac{1}{2} \tilde \mu\left(\psi^2 + \phi\square\phi\right) +
m^2\,  \phi \psi    \right\} \, .
\ee
The TME limit is one in which $\tilde \mu\to\infty$ for fixed $\mu\equiv m^2/\tilde\mu$, which implies that $m^2\to\infty$ too.  In this 
limit, $\psi$ is auxiliary and can be eliminated to yield a unitary action for $\phi$ that propagates a single mode of mass $\mu$. 
Setting $\tilde\mu=0$ gives us the EP model that is on-shell equivalent to Proca; this model should therefore propagate two modes of mass $m$, and this is clearly the case.  However, it follows from the obvious diagonalization of the action when $\tilde\mu=0$ that one 
mode  is physical and the other a ghost. 

When $\tilde\mu$ is neither zero or infinity, we may define new variables $(U,V)$ by
\be
\sqrt{\tilde\mu}\ \phi= U+V \, , \qquad \psi =\sqrt{\tilde \mu}\ V \, ,
\ee
to arrive at the action
\be\label{UVact}
S[U,V] = \int \! d^3 x \left\{ \tfrac{1}{2}\left(U\square U - V\square V \right) + m^2 \left(U+V\right)V + \tfrac{1}{2}\tilde\mu^2 V^2 \right\}\, .
\ee
We see that there is a ghost, in agreement with \cite{Deser:1999pa}.  The Hamiltonian can be diagonalized and the ghost decoupled by taking its mass to infinity,  but this just leads back to TME. Thus, the addition  of  an ECS term to a TME action, which by itself propagates a massive mode of helicity $\pm 1$,  leads to a modification of the mass of this TME mode {\it and}  the propagation of an additional massive ghost mode of helicity $\mp 1$. 

The focus of \cite{Deser:1999pa} was the ``extended topologically massive electrodynamics'' (ETME) model with action
\be
S_{ETME}[A] = \frac{1}{2}\int \! d^3 x \left\{ \tilde F^2 - \frac{1}{\mu} 
\varepsilon^{\mu\nu\rho} \tilde F_\mu \partial_\nu \tilde F_\rho  \right\}\, .
\ee
The field equation is 
\be
\Box \tilde F^\mu = \mu\, \varepsilon^{\mu\nu\rho} \partial_\nu \tilde F_\rho\, . 
\ee
Compare this to the TME field equation 
\be
\Box  A^\mu = \mu\, \varepsilon^{\mu\nu\rho} \partial_\nu  A_\rho + \partial^\mu \left(\partial\cdot A\right)\, . 
\ee
We see that the ETME equation is obtained from the TME equation {\it by replacing the gauge potential  by its field 
strength}. One might think that this implies their equivalence since a zero potential  implies a zero field strength,  and a zero field strength implies a pure-gauge potential, which is effectively zero in the context of a gauge theory such as TME. However, they are not equivalent: by setting $m^2=0$ in (\ref{UVact}), one sees that ETME  propagates a massless mode in addition to a  massive one
\cite{Deser:1999pa}. The overall sign  is such that the massive mode is a ghost but we could change the sign to make this physical, in which case this `reversed' ETME would differ from  TME  by the additional propagation of a massless ghost.  What we wish to emphasize
here is that {\it the procedure of  replacing a field by its field strength is not one that invariably leads to equivalent field equations}.  There is no contradiction with our procedure of solving subsidiary conditions for the simple reason that 
TME has no subsidiary conditions.

\section{Higher-derivative 3D gravity}\label{sec:spin2} 
\setcounter{equation}{0}

{}For spin 2,  we start by considering the following action for a tensor field $e_{\mu\nu}$ of {\it no definite symmetry}:
\be\label{eact}
S[e] =  -\frac{1}{2}\int \! d^3 x\, \left\{ e^{\mu\nu} {\cal G}_{\mu\nu}{}^{\rho\sigma} e_{\rho\sigma} + \frac{\tilde\mu}{2} \varepsilon^{\mu\tau\rho}e_\mu{}^\nu \partial_\tau e_{\rho\nu}
+ \frac{m^2}{2}\left(e^{\nu\mu}e_{\mu\nu} - e^2\right) \right\}\, ,
\ee
where $e= \eta^{\mu\nu}e_{\mu\nu}$, and  ${\cal G}$ is the Einstein operator defined in (\ref{tildehG}). Note that ${\cal G}$ is, by definition, symmetric in both index pairs; as a consequence, only the symmetric part of $e_{\mu\nu}$ enters into the 
first term in (\ref{eact}). If we write 
\be
e_{\mu\nu} = \phi_{\mu\nu} + \varepsilon_{\mu\nu\rho}\,  t^\rho \, ,
\ee
where $\phi$ is symmetric, then one finds that the field equations are equivalent to
\be\label{genspin2}
[{\cal P}(m_+){\cal P}(-m_-)]_\mu{}^\rho\, \phi_{\rho\nu} =0\, , \qquad \eta^{\mu\nu}\phi_{\mu\nu}=0\, , \qquad t_\mu=0\, .
\ee
The `auxiliary' vector $t$ appears irrelevant  but is needed for the action, as shown in  \cite{Aragone:1986hm} in the context of the self-dual  case obtained in the limit $m_-\to\infty$ for fixed $m_+$. One may verify that the Hamiltonian density associated to the action
is positive definite, which implies that the two spin 2 modes are propagated unitarily. 

As explained in the introduction, the differential subsidiary constraint  $\partial^{\nu}\phi_{\mu\nu}=0$ implied by the equations (\ref{genspin2}) (assuming $m^2\ne0$) has the general solution (\ref{tildehG}) in terms of a symmetric tensor potential $h$.   Making this substitution  in the generalized FP equations (\ref{genspin2}),  we arrive at the equations
\be\label{linGMG}
[{\cal P}(m_+){\cal P}(-m_-)]_\mu{}^\rho\, G^{\rm (lin)}_{\rho\nu} (h) =0\, , \qquad \qquad R^{\rm (lin)}(h)=0\, .
\ee
These are precisely the field equations of the linearized GMG model obtained by adding an LCS term to the parity-preserving NMG model. This guarantees on-shell equivalence of GMG to the generalized FP theory. It does not guarantee off-shell equivalence, i.e. the absence of ghosts, but this was  shown in \cite{Andringa:2009yc} following the canonical analysis used for NMG in \cite{Deser:2009hb}.

As also explained in the introduction, one may solve both the differential and the algebraic tracelessness condition by equating $\phi$ to the linearized Cotton tensor for a symmetric tensor gauge potential $h$ that is now defined up to the combination of a linearized diffeomorphism and a linearized Weyl transformation.  In this way the first-order self-dual spin 2 equations become the linearized equations of the 4th order NTMG theory \cite{Andringa:2009yc}. Here we shall use the same procedure  to convert the  second-order generalized FP equations for spin 2 into equivalent 5th order equations.  Specifically, we use (\ref{newsub}) 
in the 2nd order equations (\ref{genspin2}) to  arrive at the equivalent 5th order equation
\be
[{\cal P}(m_+){\cal P}(-m_-)]_\mu{}^\rho\, C^{\rm (lin)}_{\rho\nu} =0\, , 
\ee
which is invariant  under the gauge transformations
\be\label{gaugetrans}
h_{\mu\nu} \rightarrow  h_{\mu\nu} + \partial_{(\mu}v_{\nu)} + \eta_{\mu\nu}\,  \omega
\ee
for arbitrary vector field $v$ and scalar field $\omega$. The subsidiary conditions have now been replaced by the gauge-invariant  identities
\be
\partial^\mu C^{\rm (lin)}_{\mu\nu} \equiv 0\, , \qquad \eta^{\mu\nu} C^{\rm (lin)}_{\mu\nu} \equiv 0\, . 
\ee
The  gauge-invariant action is
\be\label{high-deriv}
S[h] = \mu\int\,d^3 x\,\left\{-\frac{1}{2}h^{\mu\nu} C^{\rm (lin)}_{\mu\nu}  -
\frac{1}{2\mu} \varepsilon^{\mu\tau\rho} h_\mu{}^\nu \partial_\tau C^{\rm (lin)}_{\rho\nu}
+\frac{1}{2m^2} h^{\mu\nu} \square C^{\rm (lin)}_{\mu\nu} \right\}\, . 
\ee
This is the quadratic approximation to the 5th order action
\be
S= \mu\int\,d^3 x \left\{- {\cal L}_{LCS} + \frac{1}{\mu} \sqrt{|g|} \, K - \frac{1}{m^2}\sqrt{|g|} R^{\mu\nu}C_{\mu\nu}\right\} \, ,
\ee
which includes (for finite $m^2$) an  ELCS term.  Note that the linearized Weyl invariance of the linear theory does not generalize to the full theory. Instead one finds a conformal covariance property, explained for NMG   in  \cite{Bergshoeff:2009hq}: a scalar  density  is conformally covariant if the Weyl variation of its integral is the integral of a function times the same scalar density. This is a property 
of the scalar density $\sqrt{|g|} \, K$ of NMG and it is also a property of the ELCS term. As both these scalar densities have no linear term, their conformal covariance implies a linearized Weyl invariance of the quadratic approximation. 

We now have a new 5th order theory for spin 2 but there is no guarantee that neither  spin 2 mode is a ghost. We shall investigate this  issue in the context of the linearized theory with quadratic action (\ref{high-deriv}). The six components of the metric perturbation $h_{\mu\nu}$ are determined up to the gauge transformation (\ref{gaugetrans})   by two functions.  We may choose the gauge such that
\be
h_{ij}=0\, , \qquad \partial_i h_{0i}=0\, ,
\ee
which means that
\be
h_{0i} = - \varepsilon^{ij} \frac{1}{\nabla^2} \partial_j \xi\, , \qquad h_{00}= \frac{1}{\nabla^2} N\, ,
\ee
for two functions $(\xi,N)$. This decomposition can be obtained from that given in  \cite{Andringa:2009yc} (which is a variation of that used in  \cite{Deser:2009hb}) by imposing the additional Weyl gauge condition $\delta^{ij} h_{ij}=0$, and the results for the Weyl-invariant Cotton tensor components are therefore identical to those given in \cite{Andringa:2009yc}. Here we go directly to the quadratic action, which becomes
 \be\label{quadaction1}
 S[\xi,N] = \frac{1}{2}\mu\int\,d^3 x \left\{- \xi N +  \frac{1}{\mu} \left(\xi\square\xi + \frac{1}{4} N^2\right) + \frac{1}{m^2} N\square \xi  \right\}\, .
 \ee
In the $m^2\to\infty$ limit for fixed $\mu$ we recover the result of \cite{Andringa:2009yc} for the NTMG model; the variable $N$ is auxiliary in this limit and its elimination yields an action that propagates unitarily a single mode of mass $\mu$.  

If we first divide by the overall factor of $\mu$ in (\ref{high-deriv}) then we may take the $\mu\to\infty$ limit for fixed $m^2$ to  get a 5th order model with the following Lagrangian: 
\be\label{analog}
L= -\frac{1}{2}h^{\mu\nu} C^{\rm (lin)}_{\mu\nu} +\frac{1}{2m^2} h^{\mu\nu} \square C^{\rm (lin)}_{\mu\nu} \, . 
\ee
By construction, this is on-shell equivalent to the standard 3D FP model for massive spin 2.  However, applying to (\ref{quadaction1}) 
the  same procedure of dividing by $\mu$ and then taking the $\mu\to\infty$ limit we arrive at the `reduced' Lagrangian 
\be
L=  \frac{1}{m^2} N \left(\Box-m^2\right)\xi\, , 
\ee
which shows that one mode is a ghost.  The 5th order model with Lagrangian (\ref{analog})  is therefore a much closer spin 2 analog of the EP model than is linearized NMG, so it could be viewed as the linearized version of an ``extended Fierz-Pauli'' theory. Alternatively, the non-linear theory with action
\be
S=  \int d^3x \left\{ {\cal L}_{LCS} + \frac{1}{m^2} \sqrt{|g|} R^{\mu\nu}C_{\mu\nu}\right\} \, ,
\ee
could be viewed as ``extended NMG'' (ENMG) theory. 

Returning to (\ref{quadaction1}) and assuming that $\mu$ is neither zero nor infinity, we  may define new variables $(U,V)$  by
\be
\xi = \frac{1}{\sqrt{2}} \left(U + V\right)\, , \qquad N = \sqrt{2}\, \tilde\mu\, V\, , 
\ee
to arrive at the equivalent action
\be\label{UVact2}
S[U,V] = \int \! d^3 x \left\{\tfrac{1}{2} \left(U \square U- V \square V\right) +
m^2 \left(U+V\right) V+ \tfrac{1}{2}\tilde\mu^2 V^2 \right\}\, .
\ee
This is {\it identical} to (\ref{UVact}), which means that the ELCS term leads to ghosts in 3D gravity in precisely the same way that the ECS term leads to ghosts in 3D electrodynamics.  We next show how this fact is made manifest in ${\cal N}=2$ 3D supergravity.

\subsection{Higher-derivative ${\cal N}=2$ supergravity}

We now aim to embed the action (\ref{high-deriv}) into one of  linearized ${\cal N}=2$ 3D supergravity. Because of the linearized Weyl invariance, we need only consider the Weyl supermultiplet, which consists of the fields $\{h_{\mu\nu}, \psi_\mu^a, A_\mu\}$,  where the (anti-commuting) RS fields $\psi_\mu^a$  ($a=1,2$) are Majorana spinors. These fields are subject to the following gauge invariances that reduce the number of independent off-shell degrees of freedom to 4 bosons and 4 fermions:
\be\label{lingaugetransf}
\delta h_{\mu\nu} = \partial_{(\mu}v_{\nu)} + \eta_{\mu\nu}\omega\,,\qquad
\delta\psi_\mu^a = \partial_\mu \zeta^a + \gamma_\mu\eta^a\,,\qquad
\delta A_\mu =  \partial_\mu \alpha\, .
\ee
The vector field $v_\mu$ is the parameter of linearized diffeomorphisms, the scalar fields $\omega$ and $\alpha$ are the Weyl and abelian gauge parameters, respectively, and  the Majorana spinor fields $\eta^a$ and $\zeta^a$ are the parameters of linearized S-supersymmetry and Q-supersymmetry, respectively.  In the linearized theory, the Q-supersymmetry gauge invariance is independent from the rigid supersymmetry transformations, which are
\begin{eqnarray}\label{lintrrules}
\delta h_{\mu\nu} &=& {\bar\epsilon}^a\gamma_{(\mu}\psi_{\nu)}^a\,,\nonumber\\%[.2truecm]
\delta\psi_\mu^a &=& - \tfrac{1}{4}  \partial_\rho h_{\sigma\mu}\, \gamma^{\rho\sigma}\epsilon^a
- \tfrac{1}{2}A_\mu\, \epsilon^{ab}\epsilon^b \,, \\
\delta A_\mu &=& \tfrac{1}{2} \epsilon^{ab} \, {\bar\epsilon}^{a}\phi_\mu^{b}\, ,
\qquad \phi_\mu^a \equiv \gamma_\nu\gamma_\mu {\cal R}_{\rm (lin)}^{\nu\,a} \nonumber\,.
\end{eqnarray}
This result agrees with  \cite{Andringa:2009yc} when account is taken of the linearized gauge invariances.  The vector-spinors $\phi_\mu^a$ are  the (dependent) S-gauge fields;  the RS field-strengths are defined as in (\ref{Cottino}). The transformations (\ref{lintrrules}) imply the following 
${\cal N}=2$ supersymmetry transformations of the conformal field strengths:
\begin{eqnarray}
\delta C_{\mu\nu}^{\rm (lin)} &=&  
\tfrac{1}{4}{\bar\epsilon}^a\gamma^{}_{\rho(\mu}\partial^\rho{\cal C}_{\nu)}^{a\,{\rm (lin)}}\,,   \nonumber\\
\delta{\cal C}^{\mu\, a}_{\rm (lin)} &=&  \gamma_\nu\epsilon^a C^{\mu\nu}_{\rm (lin)}
+\tfrac{1}{2}\varepsilon^{\mu\nu\rho}\varepsilon^{ab}\gamma_\sigma\gamma_\nu\epsilon^b\partial_\rho \tilde{F}^\sigma\,,  \\
\delta \tilde{F}^\mu &=& \tfrac{1}{2}\varepsilon^{ab}{\bar\epsilon}^a{\cal C}^{\mu\, b}_{\rm (lin)}\,.    \nonumber
\end{eqnarray}
Here,  ${\cal C}_{\rm (lin)}^{\mu\, a}$ is the doublet of ``Cottino" vector-spinors; as defined for ${\cal N}=1$ in (\ref{Cottino}).
Using these transformation laws, one may verify invariance of the action
\be\label{linaction}
S^{{\cal N}=2} = \int d^3x\,\left\{\tfrac{1}{2} h^{\mu\nu}C_{\mu\nu}^{\rm (lin)} +\tfrac{1}{2} {\bar\psi}^a_\mu{\cal C}_{\rm (lin)}^{\mu\,a}
- \tfrac{1}{2} A_\mu \tilde F^\mu\right\}\, . 
\ee
To check the invariance one needs the various Bianchi identities satisfied by the superconformal field-strengths, the tracelessness of the Cotton tensor and the `gamma tracelessness' of the Cottino tensors. In agreement with \cite{Andringa:2009yc}, we see that the CS term of 3D electrodynamics and the LCS term of 3D gravity appear in the same ${\cal N}=2$ invariant. However, the formula (\ref{linaction}) can now be used to construct a whole family of  ${\cal N}=2$ supersymmetric actions.  

For each set of fields  that transform as in (\ref{lintrrules}) modulo linearized gauge transformations of the form (\ref{lingaugetransf}),  the action formula (\ref{linaction}) yields an ${\cal N}=2$ super-invariant. For example,  one such set of fields is
\be\label{repl(1)}
h^{(1)}_{\mu\nu} =  \varepsilon_{(\mu}{}^{\rho\sigma}\partial_{|\rho} h_{\sigma|\nu)}\,, \qquad
\psi_\mu^{a\,(1)}= \varepsilon_\mu{}^{\nu\rho} \partial_\nu\psi_\rho^a\,,   \qquad
A^{(1)}_\mu = \varepsilon_\mu{}^{\nu\rho}\partial_\nu A_\rho\,.
\ee
To be more precise, these derived objects transform as in (\ref{lintrrules}) provided one ignores `pure-gauge' terms  of the form (\ref{lingaugetransf}) (which is justified because they automatically drop out of  a gauge-invariant action).  To verify 
this, we first note that  $\psi_\mu^{a\,(1)}$ transforms into $A^{(1)}_\mu$ in the required way; the candidate for $h_{\mu\nu}^{(1)}$
found this way is not manifestly symmetric but this  can be remedied by adding field-dependent Q- and S-gauge transformations.  Similarly one finds that $h_{\mu\nu}^{(1)}$ does not transform into $\psi_\mu^{a\,(1)}$ but instead that
 \bea
  \delta h_{\mu\nu}^{(1)} = \tfrac{1}{4}\bar{\epsilon}^{a}\gamma_{\mu}\left(
  \gamma^\tau\partial_\tau\psi_\nu^{a} + \varepsilon_\nu{}^{\rho\sigma} \partial_\rho
  \psi_\sigma^a\right)+(\mu\leftrightarrow \nu)\, .
 \eea
However, the first term can be rewritten as the second term up to a Q- and S-gauge transformation by making use of the identity
\be
\varepsilon_{\mu}{}^{\rho\sigma} \equiv \gamma_{\mu}\gamma^{\rho\sigma}-
2\delta_{\mu}{}^{[\rho}\gamma^{\sigma]}\, . 
\ee 
Using (\ref{repl(1)}) in the action  formula (\ref{linaction}) yields the supersymmetric invariant
\be\label{linaction(1)}
S_{(1)}^{{\cal N}=2} = \int d^3x\,\left\{\tfrac{1}{2} h_{(1)}^{\mu\nu}C_{\mu\nu}^{\rm (lin)} +\tfrac{1}{2} {\bar\psi}^{a\,(1)}_\mu{\cal C}_{\rm (lin)}^{\mu\,a}
- \tfrac{1}{2} A^{(1)}_\mu \tilde{F}^\mu\right\}\,.
\ee
This is  the ${\cal N}=2$ supersymmetrization of the curvature-squared $K$ invariant of NMG, and is again in agreement\footnote{To compare,  one should again use that the  expression $\psi_\mu^{a\,(1)}=\epsilon_\mu{}^{\nu\rho} \partial_\nu\psi_\rho^a$ occurring in the action (\ref{linaction(1)}) differs from the expression  $\gamma^\tau\partial_\tau\psi^a_\mu$ occurring in (7.9) of 
\cite{Andringa:2009yc} by a Q- and S-gauge transformation.} with the result of  \cite{Andringa:2009yc}.

At this stage we have merely confirmed the results of \cite{Andringa:2009yc} for the ${\cal N}=2$  supersymmetrization of the linearized LCS and curvature-squared terms. However, we may now go further by iterating (\ref{repl(1)}) to get a new set of fields, again up to pure-gauge terms:
\be
h^{(2)}_{\mu\nu} =   -2R_{\mu\nu}^{\rm (lin)}\,,\qquad
\psi^{a\,(2)}_\mu = \tfrac{1}{2}{\cal C}_\mu^{a\,{\rm (lin)}}\,, \qquad
A^{(2)}_\mu = \epsilon_\mu{}^{\nu\rho}\partial_\nu \tilde{F}_\rho\,.
\ee
The transformations of these fields are again of the required form, so we may use them in the action formula (\ref{linaction}) to get a new ${\cal N}=2$ invariant that includes both the ECS and the ELCS terms: 
\be\label{linaction(2)}
S_{(2)}^{{\cal N}=2} =  \int d^3x\, \left\{ - R^{\mu\nu}_{\rm (lin)} C_{\mu\nu}^{\rm (lin)}  + 
\tfrac{1}{4} \bar{\cal C}_\mu^{\rm (lin)}{}^a  {\cal C}_{\rm (lin)}^{\mu\,a}
- \tfrac{1}{2} \varepsilon^{\mu\nu\rho} \tilde F_\mu\partial_\nu \tilde F_\rho\right\}\, . 
\ee
This result implies that the ECS and ELCS terms lead to ghosts in precisely the same way since any ghosts must fill out on-shell ${\cal N}=2$ particle multiplets.

It should now be clear that we can construct, by further iteration,  any number of still higher-derivative ${\cal N}=2$ invariants. For example, the following set of fields has the required transformations
\be
h^{(3)}_{\mu\nu}=   -2C_{\mu\nu}^{\rm (lin)}\,,\qquad
\psi^{a\,(3)}_\mu =  \tfrac{1}{2}\gamma^\tau\partial_\tau {\cal C}^{a\,{\rm (lin)}}_\mu\,,\qquad
A^{(3)}_\mu = \square \tilde{F}_\mu\;.
\ee
Substitution into the action formula  yields an ${\cal N}=2$ invariant containing the $F\square F$ invariant of electrodynamics and the square of the Cotton tensor.

%%%%%%%%%%%%%%%%%%%%%%%%%%%%%%%%%%%%%%%%%%%

\section{Spin 3}\label{sec:spin3}
\setcounter{equation}{0}

Let us start from the first-order  self-dual  equations for spin 3:
\be\label{spin3SD}
{\cal P}(\mu)_\mu{}^{\sigma} \phi_{\sigma\nu\rho} =0\, , \qquad \phi_\mu\equiv \eta^{\nu\rho}\phi_{\mu\nu\rho}=0\, .  
\ee
These imply the differential subsidiary condition $\partial^\mu\phi_{\mu\nu\rho}=0$, 
which has the following general solution in terms of a 3rd rank symmetric tensor potential $h$:
\be\label{gensoln3}
\phi_{\mu\nu\rho} =  {\cal G}_{\mu\nu\rho}{}^{\alpha\beta\gamma} h_{\alpha\beta\gamma} \equiv G_{\mu\nu\rho}\, , 
\ee
where
\be\label{spin3Einstein}
 {\cal G}_{\mu\nu\rho}{}^{\alpha\beta\gamma} \equiv 
- \frac{1}{6} \varepsilon_{(\mu}{}^{\tau\alpha} \varepsilon_\nu{}^{\sigma\beta}\varepsilon_{\rho)}{}^{\eta\gamma}
\partial_\tau\partial_\sigma\partial_\eta 
\ee
is the spin 3 analog of the Einstein operator. The tensor $G_{\mu\nu\rho}$ is a spin-$3$ analog of the linearized Einstein tensor;  it is invariant under the spin 3 gauge transformation
\be\label{diff3}
h_{\mu\nu\rho} \rightarrow h_{\mu\nu\rho}+ \partial_{(\mu} \xi_{\nu\rho)}
\ee
for arbitrary symmetric tensor field $\xi$. The self-dual spin 3 equations are therefore equivalent to the 4th order equations
\be
{\cal P}(\mu)_\mu{}^{\sigma} G_{\sigma\nu\rho} =0 \, , \qquad  G_\mu\equiv \eta^{\nu\rho}G_{\mu\nu\rho}=0\, .  
\ee
The first of these equations implies the second, but only because of the lack of symmetry in its three indices; an alternative
equivalent set of  equations is
\be\label{altset}
{\cal P}(\mu)_{(\mu}{}^{\sigma} G_{\nu\rho)\sigma} =0\, ,  \qquad  G_\mu=0\, . 
\ee
By construction, these equations propagate a single spin 3 mode (and we verify this below).  They are the natural generalization to spin $3$ of the linearized TMG equations for spin $2$. We expect the action  to involve auxiliary fields, which we will not attempt to find here. However, it will obviously be possible to choose the overall sign of the action so as to ensure unitarity, and hence a  4th order unitary theory describing a single spin-$3$ mode. 

Before proceeding to other cases, we should mention that a 3rd order action for a symmetric tensor gauge field was proposed in 
\cite{Damour:1987vm} as a spin-$3$ analog of linearized TMG. Since the symmetric tensor parameter of the spin-$3$ gauge transformation was restricted to be traceless, it was possible to find a 2nd order rank-$3$ invariant tensor analogous to the Einstein tensor, and the equation proposed in \cite{Damour:1987vm} is formally the same as the first of the equations (\ref{altset}) but with 
our 3rd order rank-$3$ tensor $G$ replaced by their 2nd order rank-$3$ tensor. The resulting equation is derivable from an action, but this (3rd order) action propagates a spin 1 ghost in addition to the desired massive spin 3 mode \cite{Damour:1987vm}. As stated above, we believe that the 4th order equations (\ref{altset}) are the natural generalization to spin $3$ of the linearized TMG equations.
There is a further generalization to spin $s$ that we present in the following section.  

We now turn to the standard  FP equations for spin 3 
\be\label{genspin3}
\left(\Box -m^2\right) \phi_{\mu\nu\rho} =0\, , \qquad \partial^\mu\phi_{\mu\nu\rho}=0 \, , \qquad \eta^{\mu\nu} \phi_{\mu\nu\rho} =0\, . 
\ee
Solving the differential subsidiary constraint, as in (\ref{gensoln3}), we  arrive at the 5th order equations
\be\label{FP3gauge}
\left(\Box - m^2\right) G_{\mu\nu\rho} =0\, , \qquad G_\mu=0\, .  
\ee
One easily shows that these equations are equivalent to 
\be\label{GplusC}
G_{\mu\nu\rho} - \frac{2}{m^2} C_{\mu\nu\rho} =0\, ,  
\ee
where 
\be\label{spin3C}
C_{\mu\nu\rho} = \frac{1}{2} \Box G_{\mu\nu\rho} - \frac{3}{8}\Theta_{(\mu\nu} G_{\rho)} \, , \qquad 
\Theta_{\mu\nu} \equiv  \eta_{\mu\nu} \Box - \partial_\mu\partial_\nu\, , 
\ee
which is identically traceless: $\eta^{\nu\rho}C_{\mu\nu\rho}\equiv 0$.
A crucial property of both  $G$ and $C$ is that they take the form
\be
G_{\mu\nu\rho} = {\cal G}_{\mu\nu\rho}{}^{\alpha\beta\gamma} h_{\alpha\beta\gamma}\, , \qquad
C_{\mu\nu\rho} = {\cal C}_{\mu\nu\rho}{}^{\alpha\beta\gamma} h_{\alpha\beta\gamma}\, , 
\ee
for {\it self-adjoint}  matrix operators ${\cal G}$ and ${\cal C}$.  This means that the equation (\ref{GplusC}) follows from variation of the action
\be\label{spin3NMG}
S= \frac{1}{2}\int d^3 x\,  h^{\mu\nu\rho} \left[ {\cal G}_{\mu\nu\rho}{}^{\alpha\beta\gamma} - \frac{2}{m^2} {\cal C}_{\mu\nu\rho}{}^{\alpha\beta\gamma}\right] h_{\alpha\beta\gamma}\, . 
\ee
This is the spin 3 analog of the linearized NMG action for spin 2 and the extended Proca action for spin 1. The construction does not guarantee the absence of ghosts, so we must address this question by other methods. We shall see that  one of the two spin 3 modes is a ghost. For this reason, we will not bother to investigate the gauge-invariant formulation of the  generalized FP equations for spin 3. 

The rank-$3$ tensor $C$ defined in (\ref{spin3C}) is the spin 3 analog of the linearized Cotton tensor; it is 
 invariant under the spin 3  analog of the linearized Weyl transformation 
\be
h_{\mu\nu\rho} \rightarrow h_{\mu\nu\rho} + \eta_{(\mu\nu}\omega_{\rho)}\, , 
\ee
in addition to its invariance under the transformation (\ref{diff3}), and it satisfies the identities 
\be
\partial^\mu C_{\mu\nu\rho} \equiv 0 \, , \qquad \eta^{\nu\rho} C_{\mu\nu\rho} \equiv 0\, . 
\ee
An alternative expression is 
\be\label{altC}
C_{\mu\nu\rho} = \frac{1}{2} \varepsilon_{(\mu}{}^{\tau\alpha} \varepsilon_\nu{}^{\sigma\beta}
\partial_{|\tau}\partial_{\sigma|}S_{\rho)\alpha\beta}\, ,  \qquad 
S_{\mu\nu\rho} \equiv  G_{\mu\nu\rho} - \frac{3}{4} \eta_{(\mu\nu} G_{\rho)}\, . 
\ee
The rank-$3$ tensor $S$ is  the spin $3$ analog of the Schouten tensor. Note the explicit symmetrization that is needed in this expression for $C$,  in contrast to the analogous formula for spin 2 in (\ref{CotSchout}).  The 5th order action
\be
S_3^{\rm conf} = \int d^3 x \, h^{\mu\nu\rho} C_{\mu\nu\rho}
\ee
is the ``conformal spin 3'' action presented in  \cite{Pope:1989vj}, which has no propagating modes.

As in the spin 2 case, we may also solve  {\it both} the differential and the algebraic subsidiary conditions by writing
\be\label{both}
\phi_{\mu\nu\rho} =   C_{\mu\nu\rho}\, . 
\ee
We thus find that the standard spin-$3$ equations are equivalent to the  7th order equation
\be
\left(\Box -m^2\right) C_{\mu\nu\rho} =0\, . 
\ee
This is derivable from the action
\be\label{spin3Cac}
S= \frac{1}{2}\int d^3 x\,  h^{\mu\nu\rho} \left(\Box -m^2\right) C_{\mu\nu\rho}\, . 
\ee
We shall show below that this propagates one of the two spin-$3$ modes as a ghost, so this spin $3$ model is 
analogous to the EP  model for spin $1$. 

The problem of ghosts should not  arise in the self-dual limit, in which only one mode is propagated. In this case, we use 
 (\ref{both}) in (\ref{spin3SD}) to arrive at the equivalent equation 
\be\label{ordersix}
{\cal P}(\mu)_\mu{}^{\sigma} C_{\sigma\nu\rho} =0\, . 
\ee
This is a 6th order  differential equation for the gauge  potential $h$, invariant under the gauge transformation
\be
\delta h_{\mu\nu\rho} = \partial_{(\mu} \xi_{\nu\rho)} + \eta_{(\mu\nu}\omega_{\rho)}\, , 
\ee
for arbitrary symmetric tensor parameter $\xi$ and vector parameter $\omega$.  It should be appreciated that this  gauge potential is not the same as the one in (\ref{gensoln3}); their dimensions are different. The equation (\ref{ordersix})  is derivable from the 6th order gauge-invariant action 
\be\label{gaugeSD3}
S = \frac{1}{2}\int d^3 x\, h^{\mu\nu\rho} {\cal P}(\mu)_\mu{}^{\sigma} C_{\sigma\nu\rho}\, . 
\ee
We shall verify below that this action propagates, unitarily, a single (spin 3) mode. This is the spin $3$ analog of NTMG.

\subsection{Canonical analysis}

We may choose the gauge
 \be
\partial_i h_{i\nu\rho} =0\, . 
\ee
In this gauge we may write
\begin{eqnarray}
h_{000} &=& \frac{1}{\left(-\nabla^2\right)^{\frac{3}{2}} }\phi_0\, , \qquad 
h_{00i} = \frac{1}{\left(-\nabla^2\right)^{\frac{3}{2}}}\hat \partial_i \phi_1\, , \nonumber \\
h_{0ij} &=& \frac{1}{\left(-\nabla^2\right)^{\frac{3}{2}}} \hat\partial_i \hat\partial_j\phi_2\, , \qquad 
h_{ijk} = \frac{1}{\left(-\nabla^2\right)^{\frac{3}{2}}}\hat\partial_i \hat\partial_j \hat\partial_k \phi_3\, , 
\end{eqnarray}
where we recall that, by definition,
\be
\hat\partial_i  = \varepsilon^{ij} \partial_j \, . 
\ee
Note the following useful identities
\be
\hat\partial_i \hat\partial_j   \equiv \delta_{ij} \nabla^2 - \partial_i\partial_j \, , \qquad \hat\partial_i  \partial_i \equiv 0\, . 
\ee
A computation of the components of the tensor $G_{\mu\nu\rho}$ yields the result
\begin{eqnarray}
G_{000} &=&  \frac{\left(-\nabla^2\right)^{\frac{3}{2}}}{6}  \phi_3 \, , \qquad 
G_{00i} =   -\frac{ \left(-\nabla^2\right)^{\frac{1}{2}}}{6} \left[ \hat\partial_i \phi_2 + \partial_i \dot\phi_3\right] \, , \nonumber \\
G_{0ij} &=& \frac{1}{6  \left(-\nabla^2\right)^{\frac{1}{2}}}\left[ \hat\partial_i\hat\partial_j \phi_1 + 2\hat\partial_{(i}\partial_{j)} \dot\phi_2
+   \partial_i\partial_j \ddot\phi_3 \right]  \, , \\
G_{ijk} &=& -\frac{1}{6  \left(-\nabla^2\right)^{\frac{3}{2}} } \left[ \hat\partial_i\hat\partial_j\hat\partial_k \phi_0 
+ 3  \hat\partial_{(i}\hat\partial_j\partial_{k)} \dot \phi_1 + 3  \hat\partial_{(i}\partial_j\partial_{k)} \ddot \phi_2
+  \partial_i\partial_j\partial_k \partial_t^3 \phi_3\right]\, . \nonumber 
\end{eqnarray}
{}From this result we find that the components of $G_\mu$ are 
\begin{eqnarray}
G_0 &=& \frac{ \left(-\nabla^2\right)^{\frac{1}{2}}}{6} \left(\Box\phi_3 - \phi_1\right)\, ,  \nonumber\\
G_i  &=& -\frac{1}{6  \left(-\nabla^2\right)^{\frac{1}{2}}}\left[ \hat\partial_i \left(\Box\phi_2 -\phi_0\right) 
+\partial_i \left(\Box\dot\phi_3 -\dot\phi_1\right) \right]\, . 
\end{eqnarray}
Using these results, one may show that the equations (\ref{altset}) are equivalent to 
\be
\phi_0 = -m\phi_1= m^2\phi_2 = -m^3 \phi_3\, , \qquad \left(\Box -m^2\right)\phi_3=0\, , 
\ee
which confirms the propagation of a single mode of mass $m$.

Moving on to the  components of $C_{\mu\nu\rho}$, we similarly find the result
\begin{eqnarray}
C_{000} &=& \frac{\left(-\nabla^2\right)^{\frac{3}{2}}}{16}  \psi_1 \, , \qquad 
C_{00i} =  -\frac{ \left(-\nabla^2\right)^{\frac{1}{2}}}{48} \left[ \hat\partial_i \psi_0 + 3 \partial_i \dot\psi_1\right]\, , \nonumber \\
C_{0ij} &=& \frac{1}{16  \left(-\nabla^2\right)^{\frac{1}{2}}} 
\left[ \hat\partial_i\hat\partial_j \Box\psi_1 + \frac{2}{3} \partial_{(i}\hat\partial_{j)} \dot\psi_0
+ \partial_i\partial_j \ddot\psi_1\right] \, ,  \\
C_{ijk} &=&- \frac{1}{48 \left(-\nabla^2\right)^{\frac{3}{2}}} \left[ \hat\partial_i \hat\partial_j \hat\partial_k \Box\psi_0 
+9 \hat\partial_{(i }\hat\partial_j\partial_{k )} \Box\dot\psi_1 
 + 3 \hat\partial_{(i }\partial_j\partial_{k )}\ddot\psi_0 + 3 \partial_i\partial_j\partial_k \, \partial_t^3 \psi_i \right] \, , 
 \nonumber
\end{eqnarray}
where
\be
\psi_0 = \phi_0 + 3\Box \phi_2\, , \qquad \psi_1 = \phi_1 + \frac{1}{3} \Box\phi_3\, . 
\ee
We see that $C_{\mu\nu\rho}$ depends on only two linearly independent combinations of the four variables $(\phi_0,\phi_1,\phi_2,\phi_3)$. 

To analyse the Lagrangian (\ref{spin3NMG})  we need the following results:
\begin{eqnarray}
h^{\mu\nu\rho} G_{\mu\nu\rho} &=&  -\phi_1\phi_2 -\frac{1}{3}\phi_0\phi_3 \  =\ - \psi_1 \phi_2
 - \frac{1}{3} \psi_0 \phi_3  + \frac{4}{3} \phi_2 \Box\phi_3\, , \nonumber\\
h^{\mu\nu\rho} C_{\mu\nu\rho} &=&  -\frac{1}{8} \psi_0\psi_1\, .
\end{eqnarray}
We now find that
\begin{eqnarray}
h^{\mu\nu\rho} \left[ G_{\mu\nu\rho} - \frac{2}{m^2} C_{\mu\nu\rho}\right] 
= \left(\frac{1}{4m^2} \psi_0 - \phi_2\right)\left(\psi_1 - \frac{4m^2}{3}\phi_3\right)+ \frac{4}{3}\phi_2\left(\Box -m^2\right)\phi_3\, . 
\end{eqnarray}
The first term is zero on using the $(\psi_0,\psi_1)$ equations. This leaves the second term, which propagates two modes of mass $m$ but with one of the two being a ghost, as stated earlier.  We see that the on-shell equivalence to the spin-$3$ FP theory does not extend to an off-shell equivalence.

We now turn to the higher-derivative spin 3 theory with action (\ref{spin3Cac}). The Lagrangian depends only on the variables 
$(\psi_0,\psi_1)$: 
\be
L= - \frac{1}{16} \psi_0 \left(\Box -m^2\right)\psi_1 \; .
\ee
We thus confirm that two modes of mass $m$ are propagated, but one is a ghost, as stated earlier.

Finally, we consider  the action (\ref{gaugeSD3}). We may write its Lagrangian as
\be
L= h^{\mu\nu\rho}\left[ C_{\mu\nu\rho} + \frac{1}{\mu} J_{\mu\nu\rho}\right] \, , \qquad 
J_{\mu\nu\rho} \equiv   \varepsilon_{(\mu}{}^{\tau\lambda}\partial_{|\tau} C_{\lambda |\nu\rho)} \, . 
\ee
To analyse this Lagrangian, we compute
\be
h^{\mu\nu\rho} J_{\mu\nu\rho} = \psi_0\,  \hat\partial_i C_{i00} + 3 \psi_1\,  \Box C_{000} =
 \frac{1}{48} \psi_0^2 + \frac{3}{16} \psi_1\Box\psi_1 \, . 
\ee
The first equality is obtained by using the identities satisfied by $C_{\mu\nu\rho}$. Using this result we have
\be
h^{\mu\nu\rho}\left[ C_{\mu\nu\rho} + \frac{1}{\mu} J_{\mu\nu\rho}\right] = \frac{1}{48} \left(\psi_0 -3\mu\psi_1\right)^2 
+ \frac{3}{16\mu} \psi_1\left(\Box-\mu^2\right)\psi_1 \, . 
\ee
Eliminating the auxiliary field $\psi_0$ removes the first term. What is left is a unitary Lagrangian for $\psi_1$ that propagates a single mode of mass $\mu$.  We know that this has spin $3$ so (\ref{gaugeSD3}) is  a  {\it unitary 6th order gauge-invariant action for a single spin 3 mode}.

%%%%%%%%%%%%%%%%%%%%%%%%%%%%%%%%%%%%%%%%%%%%
\section{Discussion}\label{sec:discussion}
\setcounter{equation}{0}

Relativistic field equations that are higher than second order in derivatives are typically consigned to the class of  ``higher-order'' equations exhibiting unphysical behaviour, associated classically with negative kinetic energies and quantum mechanically with negative norm states, i.e.~ghosts. However, it has been clear for some time, from the example of  the third-order ``topologically massive gravity'' (TMG) \cite{Deser:1981wh}, that  three dimensional (3D)  spacetime allows exceptions, and the recently discovered ``new massive gravity'' (NMG), and its generalization to ``general massive gravity'' (GMG) which includes both TMG and NMG as special cases, has shown that there are even fourth-order equations that are free of the bad features usually associated with ``higher derivatives'' \cite{Bergshoeff:2009hq}.  These gravitational theories are unusual in that they are generally covariant but describe {\it massive} spin 2 gravitons. 

It was shown in \cite{Andringa:2009yc} that the linearized versions of these `massive gravity'  theories can be found systematically by solving the differential subsidiary condition in standard second (or first) order `non-gauge'  massive spin-$2$ field theories of Fierz-Pauli (FP) type, and a ``new topologically massive gravity'' (NTMG) was also found this way.  Here we have developed further this method, starting with a systematic presentation of the `generalized' FP model for massive spin $s$ and its ``self-dual'' limit. We used the simpler application to spin 1 to illustrate the method, and this led us to a discussion of models related to the  ``extended topologically massive electrodynamics'' (ETME) of \cite{Deser:1999pa}.  Because the method only establishes equivalence of linear field equations, there is no guarantee of off-shell equivalence,  which amounts to the same as unitarity. This has to be checked by some other method, and a convenient one is the canonical decomposition method used in \cite{Deser:1999pa} to show that ETME has ghosts. This method was also used in \cite{Deser:2009hb} to verify the unitarity of NMG and in \cite{Andringa:2009yc}  to prove unitarity of  (super)NMG; we have made extensive further use of this method here. 

Application of these methods for spin 2 led us to a novel 5th order spin 2 theory with a natural extension to an interacting 
generally covariant massive gravity theory. The 5th order interaction term is found from the contraction of the Ricci and Cotton 
tensors  of the 3-metric. This term can be viewed as a type of Chern-Simons (CS) term and as a gravitational analog of  the ``extended
CS''  (ECS) term of ETME. In fact, this is more than an analogy: we have shown that both  ECS and ELCS terms are part of the same
${\cal N}=2$ supersymmetric invariant, which also includes the `square' of the linearized Cottino tensor-spinor.  As any ghost modes must form particle supermultiplets in this context, both ECS and ELCS terms lead to ghosts in precisely the same way, as we have explicitly verified. Although it might be considered disappointing that the new model for spin 2
fails to be unitary, we prefer to see this result as an indication of the uniqueness of GMG; there is a limit to higher derivatives if one insists on unitarity and it seems as though this limit is saturated by GMG. 

In the spin 3 case, we found a 4th order generalization of the  spin-$2$ equations of linearized TMG.  Obviously, these differ from the
3rd order equations, with a weaker gauge invariance, that were  proposed in \cite{Damour:1987vm} as a generalization of TMG; in particular, our 4th order equations actually do propagate only a single mode of spin $3$.  We expect the covariant action to require auxiliary fields but if an action exists it will be possible to ensure unitarity by a choice of sign.  In contrast, we found that the natural spin-$3$ generalization  of linearized NMG, which is of 5th order,   has ghosts.   However, we were able to find a unitary gauge-invariant 
6th order  action for a single spin 3 mode that is equivalent to the first-order self-dual theory for spin 3. This is a spin 3 analog of 
the NTMG model for spin 2 found in \cite{Andringa:2009yc,Dalmazi:2009pm}; as in that case one proceeds by finding the general simultaneous solution of both differential and algebraic subsidiary conditions. For spin 2 this is done by setting the FP field equal to a linearized Cotton tensor  for a symmetric tensor potential, and the spin 3 case  involves a rank-$3$ tensor generalization of this tensor.  
It seems likely that there is a spin-$s$ generalization of this construction: there is a spin-$s$ generalization of the linearized Cotton tensor that involves $2s-1$ derivatives \cite{Pope:1989vj}, so we expect the self-dual spin-$s$ theory to be equivalent to a conformal-type gauge theory of order $2s$. It is then natural to conjecture that $2s$ is the maximum number of derivatives consistent with unitarity for spin $s$.

For any free field theory that propagates a single mode, one may always choose the overall sign of the action (assuming that there is one) such that this one mode is physical rather than a ghost. For this reason, unitarity is not really an issue in applications of our method that start from a self-dual  model. It is only when there is more than one propagating mode that there is a potential problem: if there are modes with opposite sign kinetic energy then unitarity is lost for any choice of overall sign. The simplest example  is the ``extended Proca'' (EP) gauge theory which is  equivalent to Proca on-shell but not off-shell  because one of the two spin 1 modes it propagates 
is a ghost. One plausible explanation for this feature is that the EP vector potential has opposite parity to the vector field of the Proca theory. In fact, one may see by inspection that in all cases considered here, for
arbitrary $s$, ghosts arise in a model propagating both helicities if and only if the parity of the FP
non-gauge field differs from that of the gauge potential. This leads us to conjecture that this is a general feature. In particular, we conjecture that the gauge theory will be unitary whenever the parity of the gauge field is the same as the original FP field.  If true, this conjecture would nicely explain the unitarity of GMG, and it could simplify the construction of unitary gauge theories for higher spin. 

It is straightforward to take the first step towards a higher-spin generalization of our results. The general solution of the spin-$s$   differential subsidiary condition is
\be\label{einstein-s}
\varphi_{\mu_1\cdots\mu_s} =  - \frac{1}{s!} \varepsilon_{\mu_1}{}^{\tau_1\nu_1}\cdots 
\varepsilon_{\mu_s}{}^{\tau_s\nu_s}\partial_{\tau_1}\cdots\partial_{\tau_s} h_{\nu_1\cdots \nu_s} 
\equiv G_{\mu_1\cdots\mu_s}  \, . 
\ee
The rank-$s$ tensor $G$ is invariant under the spin-$s$ gauge transformation
\be
h_{\mu_1\cdots \mu_s}  \to h_{\mu_1\cdots \mu_s}  + \partial_{(\mu_1} \xi_{\mu_2\cdots\mu_s)}
\ee
for arbitrary rank-$(s-1)$ symmetric tensor $\xi$, and it satisfies the Bianchi-type identity
\be
\partial^\nu G_{\nu\mu_1\cdots \mu_{s-1}} \equiv 0\, , 
\ee
which replaces the differential subsidiary condition. For example, the spin-$s$ self-dual equations for a single mode of spin $s$ are equivalent to the equations
\be
{\cal P}\left(\mu\right)_{\mu_1}{}^\rho G_{\rho\mu_2\cdots\mu_s} =0\, , \qquad G_{\mu_1\cdots \mu_{s-2}} \equiv \eta^{\nu\rho}G_{\nu\rho\mu_1\cdots \mu_{s-2}} =0 
\ee
This equations generalize to spin $s$ the equations of linearized TMG.  

The linearized NMG equations can be similarly generalized. For example, the FP equations for spin 4 are equivalent to 
the 6th order equations
\be\label{spin4eqs}
\left(\Box -m^2\right)G_{\mu\nu\rho\sigma} =0 \, , \qquad \eta^{\mu\nu}G_{\mu\nu\rho\sigma}=0\, . 
\ee
This case is of interest in light of our  unitarity conjecture because the parity of $h$ is the same as the parity of $\phi$ for all {\it even integer} spins, and the spin 4 case is the first instance beyond spin 2. Thus,  we expect there to exist a {\it unitary} 6th order parity-preserving action for  spin 4. More generally, we expect  there to exist for any even positive integer $s$ a unitary parity-preserving 
action of order $s+2$ that propagates both helicity modes of spin $s$.  The problem is that it is far from clear how to find an action for these equations when $s>2$; we would guess that it exists but involves auxiliary fields.  

Other possible generalizations are to theories of all higher spins, integer or half integer, as considered for 3D in 
\cite{Blencowe:1988gj,Fradkin:1989xt,Pope:1989vj}, and to fractional spin along the lines of  \cite{Plyushchay:1990cv,Jackiw:1990ka}.  
It is also possible that there is some generalization to {\it higher dimension} but 3D is special in that the index structure of the 
independent field  is preserved in the process of solving subsidiary conditions. The 4D Proca  case illustrates the point: solving the differential subsidiary condition leads to the equation $\left(\Box -m^2\right)\tilde F^\mu=0$,  where now $\tilde F^\mu=\varepsilon^{\mu\nu\rho\sigma}\partial_\nu A_{\rho\sigma}$ for antisymmetric tensor potential $A$. There is no action for $A$ alone that yields this equation because there is no longer a match between the index structure of the potential and its equation of motion.

\subsection*{Acknowledgments}

We acknowledge helpful discussions with Nicolas Boulanger, Joaquim Gomis, Roman Jackiw, 
Luca Mezincescu, Mees de Roo and Jan Rosseel. 
PKT is supported by an EPSRC Senior Fellowship. 
The work of OH is supported by the DFG -- The German Science Foundation
and in part by funds provided by the U.S. Department of Energy (DoE) under
the cooperative research agreement DE-FG02-05ER41360.

%\begin{appendix}
%\renewcommand{\theequation}{\Alph{section}.\arabic{equation}}

%\section{...} \setcounter{equation}{0}

%\end{appendix}

\end{document}